\documentstyle[aps,aps10]{revtex}
\input epsf

\begin{document}                
\title{OPTICAL MICROSPHERE RESONATORS:
OPTIMAL COUPLING TO HIGH-Q WHISPERING-GALLERY MODES}

\author{M.L.Gorodetsky and V.S.Ilchenko}
\address{Moscow State University, 119899, Moscow, Russia}

\maketitle

\begin{abstract}
{
A general model is presented for coupling of high-$Q$ whispering-gallery
modes in optical microsphere resonators with coupler devices possessing
discrete and continuous spectrum of propagating modes. By contrast to
conventional high-Q optical cavities, in microspheres independence of
high intrinsic quality-factor and controllable parameters of coupling via
evanescent field offer variety of regimes earlier available in RF devices.
The theory is applied to the earlier-reported  data on different types of
couplers to microsphere resonators and  complemented by experimental
demonstration of enhanced coupling efficiency (about 80\%) and variable
loading regimes with $Q>10^8$  fused silica microspheres.
}
\end{abstract}


\section{INTRODUCTION}

High-$Q$ optical microsphere resonators currently attract growing interest
in experimental cavity QED \cite{prior,whispat,laser}, measurement science,
\cite{Byer,Arnold1} frequency stabilization and other photonics applications
\cite{Velich,repeat}. Stemming from extensive studies of
Mie resonances in microdroplets of aerosols \cite{Chang}(observed via elastic and
inelastic scattering of free-space beams), further studies of
laboratory-fabricated solid-state microspheres
focus on the properties and applications of highly-confined
whispering-gallery (WG) modes. The modes of this type possess negligible
electrodynamically-defined radiative losses (the corresponding radiative
quality factors $Q_{rad} > 10^{20}$ and higher), and are not accessible by
free-space beams and therefore require employment of near-field coupler
devices. By present moment, in addition
to the well known prism coupler with frustrated total internal reflection
\cite{prism,precess}, demonstrated coupler devices
include sidepolished fiber coupler \cite{Arnold1,Arnold2,Eroded}
and fiber taper \cite{taper}. The principle of all these devices is based on providing
efficient energy transfer to the resonant circular total-internal-reflection-guided
wave in the resonator (representing WG mode) through the evanescent field of
a guided wave or a TIR spot in the coupler.

It is evident a priori that efficient coupling maybe expected
upon fulfillment of two main conditions: 1)phase synchronism and
2)significant overlap of the two waves modelling the WG mode and
coupler mode respectively. Although reasonable coupling
efficiency has been demonstrated with three types of devices
(up to few tens of percents of input beam energy absorbed in a
mode upon resonance), no systematic theoretical approach
has been developed to quantify the performance of coupler
devices. It still remained unclear whether it is possible at
all and what are conditions to provide complete exchange of
energy between a propagating mode in a coupler device and the given
whispering gallery mode in high-$Q$ microsphere. Answers to these
questions are of critical importance for photonics applications
and also for the proposed cavity QED experiments with microspheres.

In this paper, we present a general approach to describe the
near-field coupling of high-$Q$ whispering-gallery mode to a
propagating mode in dielectric prism, slab or waveguide structure.
Theoretical results present a complete description and give a
recipe to obtain optimal coupling with existing devices. We
emphasize the importance of the introduced loading quality-factor
parameter $Q_c$ and its relation with the intrinsic $Q_0$-factor of WG
modes as crucial condition to obtain optimal coupling. Theoretical
consideration is complemented by experimental tests of variable loading regimes
and demonstration of improved  coupling efficiency with prism coupler.

\section{GENERAL CONSIDERATIONS}

Let us examine excitation of a single high-$Q$ whispering gallery
mode with high quality-factor by ($N$) travelling modes in an evanescent wave
coupler. This coupler can either have infinite number of spatial modes
($N=\infty$, as in prism coupler \cite{prism,precess} and slab) or only one mode ($N=1$,
as in tapered fiber\cite{taper} and integrated channel waveguide).

We shall start with simple description of the system using lumped parameters and
quasigeometrical approximation.

Let $A_0(t)$ be the amplitude of a circulating mode of total internal
reflection in the resonator (see Fig.1) to model the whispering-gallery
mode. Let the pump power be distributed in the coupler between its
modes so that $B_k(t)$ is the amplitude of mode $k$ ($1\leq k\leq N$)
and $B^{in}(t)$ is the slow varying amplitude with
$\sum|B^{in}_k(t)|^2 =|B^{in}(t)|^2$ equal to total pump power. Let us
assume for simplicity that coupling between different modes is absent
without the resonator.

Assuming that the coupling zone is much smaller than the diameter $D$ of
the resonator, we can introduce local real amplitude coefficients of
transmittance $T_k$ to describe the coupling of the resonator with all
modes of the coupler (either guided or leakage ones) and the
internal reflectivity coefficient $R$. We shall denote arrays of
transmittance coefficients and amplitudes as vectors $\bf T$ and $\bf B$
respectively.
If the quality-factor of the resonator mode is high enough, then a single
circle gives only small contribution to the mode buildup and
therefore $1-R\ll 1$.
In this case (neglecting for simplicity absorption and scattering losses
in the coupler $R_k^2=1-T_k^2$) we obtain
\begin{equation}
R=\prod R_k=1-{\sum{T_k^2}/2}=1-T^2/2
\end{equation}
The equation for the mode of the resonator will be the
following:
\begin{eqnarray}
A_0(t)&=&i  \sum{T_k B_k^{in}(t)} + R A_0(t-\tau_0) \exp{[i2\pi n_s L/\lambda-\alpha L /2]}\nonumber\\
B_k^{out}(t)&=& R_k B_k^{in}(t)+i T_k A_0(t).
\label{Fabry}
\end{eqnarray}
where $\tau_0 = n_s L/c$ is the circulation time for the mode travelling
inside the sphere, $L\simeq 2\pi a$ is approximately equal to the
circumference of the sphere, $\lambda$ is the wavelength, $n_s$ is the
refraction index, $c$ is the speed of light and $\alpha$ is the linear
attenuation in the resonator caused by scattering, absorption and radiation.

In the above representation the microsphere is equivalent to a ring resonator
formed by mirrors with transmittances $T_k$ and filled with lossy medium,
or, in case of single-mode coupling, as  pointed in \cite{Arnold2}, to a
Fabry-Perot resonator of the length $L/2$ with totally reflecting rear
mirror.

If propagation losses are small, then near the resonance frequency
$\omega_0=2\pi c/\lambda_0$, $n_sL=m\lambda_0$, where $m$ is integer,
by expanding $A_0(t-\tau_0)=A_0(t)-\tau_0\/dA_0/dt$
from (\ref{Fabry}) we obtain:
\begin{equation}
{dA_0\over dt}+(\delta_c+\delta_0+i\Delta \omega)A_0=i C B^{in},
\label{oscillator}
\end{equation}
where
\begin{equation}
\delta_0={\alpha c\over 2n_s};\;\;\;
\delta_c={1-R\over R\tau_0}={T^2\over 2\tau_0};\;\;\;
C={T \it \Gamma\over \tau_0}.
\label{oscparams}
\end{equation}

We introduce here another important coefficient:
\begin{equation}
{\it\Gamma}={{\bf T}{\bf B}^{in}\over  T  B^{in}}.
\end{equation}
This coefficient ${\it\Gamma}\leq1$ describes mode matching and shows how close
the field in the couplers matches the near field of resonator mode.

The term $\delta_0$ originates from intrinsic quality factor
$Q_0=2\pi n_s/\alpha \lambda$ while $\delta_c$ describes loading i.e.
mode energy escape to all modes of the coupler. Hereafter we shall mark all
values associated with coupler by index `$c$' and values associated with
microsphere by index `$s$'.

Equation (\ref{oscillator}) is a classical equation for the amplitude
of the resonator pumped by harmonical field.

As will be shown below, coefficients $T_k$ can be calculated as normalized
overlap integrals of the fields of the microsphere mode and modes of
the coupler. The difference from Fabry-Perot resonator is that for the
microsphere, coefficients $T_k$ are not fixed parameters but instead, strongly
depend on geometry of coupling (e.g. exponentially on the value of the gap
between microsphere and coupler) and are therefore in hands of experimentalist.
As we already emphasized
in \cite{prism}, it is the controllable relation between $\delta_0$ and
$\delta_c$ that defines coupling efficiency upon given configuration
(accounting both for mode overlap and synchronism and optimized loading to
provide energy exchange between resonator and coupler).
Stationary solution for  (\ref{oscillator}) has the typical form:
\begin{equation}
A_0={i 2\delta_c B^{in}\over \delta_0+\delta_c+i\Delta\omega}{{\it\Gamma}\over T}=
{i {\it\Gamma} B^{in}\over \delta_0+\delta_c+i\Delta\omega}\sqrt{2\delta_c\over \tau_0}
\label{in}
\end{equation}
Field amplitude in the resonator will be maximal at $\delta_c=\delta_0$
(intrinsic quality-factor equals the loaded $Q$).
The output stationary amplitudes are
\begin{equation}
{\bf B}^{out}={\bf B}^{in}-B^{in}{2\delta_c {\it\Gamma}\over
\delta_0+\delta_c+i\Delta\omega}{{\bf T}\over T}
\label{out1}
\end{equation}
and total output intensity in this case has lorentzian shape:
\begin{equation}
(B^{out})^2=(B^{in})^2\left(1-{4\delta_c\delta_0 {\it\Gamma}^2\over
(\delta_0+\delta_c)^2+(\Delta\omega)^2}\right)
\label{out2}
\end{equation}

It can be easily seen from this equation that the output signal
can be considered as the result of interference of the input
and the "re-emission" from the resonator. Note that mode
distribution of the second term in (\ref{out1}) (resonator mode
emission pattern) does not depend on the input distribution.

The most important case of (\ref{out1}) is the regime of ideal
matching (${\it\Gamma}=1$), obtained with ${\bf B}^{in}/B^{in}={\bf T}/ T$
when the fraction of the input power fed into the resonator mode is maximal.
(Single-mode coupler is always "mode-matched".) In this case,
provided $\delta_c=\delta_0$, output intensity turns to zero i.e.
the entire input power is lost inside the resonator. This regime
is usually called critical coupling.
Sometimes coupling is characterized by the fractional depth $K$ of the
resonance dip in intensity transmittance observed upon varying the
frequency of exciting wave in the coupler; from (\ref{out2})
$K$ can be expressed as follows
\begin{eqnarray}
K={4 Q_0 Q_c{\it\Gamma}^2\over (Q_0+Q_c)^2}={4Q{\it\Gamma}^2\over Q_0+Q_c}\nonumber\\
{1\over Q}={2\delta_0\over\omega}+{2\delta_c\over\omega}={1\over Q_0}+{1\over Q_c}
\label{kcoupl}
\end{eqnarray}
In case of critical coupling $K=1$ (100\%). In case of nonideal matching, critical coupling
may be observed until $2{\it\Gamma}^2>1$ (partial matching) if the output is mode-filtered
to pick up only part of the coupler modes. In this case leakage into other modes may be
considered as additional internal losses, and critical coupling is
obtained with lower loaded quality factor when
$\delta_c=\delta_0/(2{\it\Gamma}^2-1)$. If $\delta_c\gg\delta_0$ (overcoupling)
then for matched coupling the output wave in resonance has the sign opposite
to that out of resonance i.e. the resonator shifts phase by $\pi$. It is
appropriate to note here that in traditional high-Q optical resonators
comprised of mirrors, the quality-factor is limited by the
mirrors' finesse i.e. by loading. With microspheres, the situation is
opposite, and the primary role belongs to the intrinsic quality-factor.

\section{DIRECTIONAL COUPLER APPROACH}

The goal of this section is to determine parameters of the coupler-resonator system from the
electrodynamical point of view. In the recent paper \cite{alove} by D.R.Rowland and
J.D.Love, famous for their popular book on the theory of optical waveguides
\cite{blove}, the problem of coupling with whispering-gallery modes is
addressed on the basis of the model of distributed coupling between a travelling
surface mode in cylindrical resonator and a given mode in a planar (slab)
waveguide. In this approach, the coupling problem leads to the necessity
to solve a system of differential equations, which in our designations looks
as follows:
\begin{eqnarray}
{dA_0\over dz} &=& i\Delta\beta_0(z)A_0+ i C_k(z) \exp{[i(\beta_k-\beta_0)z]} B_k\nonumber\\
{dB_k\over dz} &=& i\Delta\beta_k(z)B_k+i C_k(z) \exp{[-i(\beta_k-\beta_0)z]} A_0,
\label{CME}
\end{eqnarray}
Coefficients $\Delta\beta_0(z)$ and $\Delta\beta_k(z)$ (describing perturbation of
wave numbers $\beta_0$ and $\beta_k$ of modes of the resonator and the coupler)
and distributed coupling coefficients $c_k$ can be calculated explicitly
as field cross-section integrals (see \cite{alove} and references therein).

\begin{eqnarray}
\Delta\beta_0={\omega (n_c^2 - 1)\over 8 \pi}{\int\limits_{\bf C} |{\bf e}_0|^2 ds};\;\;\;\;\;
\Delta\beta_k(z)={\omega (n_s^2 - 1)\over 8 \pi}{\int\limits_{\bf S} |{\bf e}_k|^2 ds};\nonumber\\
C_k^2={\omega^2 \over 64 \pi^2}{\int\limits_{\bf C}(n_c^2 - 1) {\bf e}_k^*{\bf e}_0 ds}{\int\limits_{\bf S}(n_s^2 - 1){\bf e}_0^*{\bf e}_k ds}
\label{Parms}
\end{eqnarray}
Here ${\bf e}_0$ and ${\bf e}_k$ are equivalent waveguide modes
of the resonator and of the coupler respectively, normalized vs. power; the
integration is done over cross-sections. Indexes $\bf S$ and $\bf C$ denote
that the integration is done inside the microsphere and coupler respectively.
In principle, conservation of energy requires that the two integrals
in expression for $C_k^2$ be equal and this is frequently postulated.
However, in common approximation that we also use here the above equality
is secured only
for phase-matched or identical waveguides, while in the opposite case the
dependence of the two integrals on the gap is different. Nevertheless, to
provide efficient coupling, this equality must be satisfied.

Parameters (\ref{Parms}) are nonzero only in the coupling zone. It may seem
that the coupler transmission matrix (CTM) and, subsequently, the above
introduced lumped $T_k$ coefficients can be found from equations (\ref{CME}).
However, analytical derivation of the output field amplitudes cannot be
found from (\ref{CME}) with exception of few simple cases. It was perhaps due
to this fact that the authors of \cite{alove} presented only numerical solution
for their particular case. Moreover, in general case CTM is a complex
2x2 matrix and cannot be characterized by one real parameter.

Fortunately, the situation is more favorable for optical microsphere resonators
with high loaded quality-factor $Q_c=\omega/2\delta_c$, when $T_k\ll 1$.
Indeed, from (\ref{oscparams}) it follows that $T^2=m/Q$. In a
fused silica resonator with the diameter $140\mu m$ ($m\simeq1000$) and
heavily loaded $Q\simeq10^7$ (intrinsic $Q$ - factor can be of the
order of $10^{10}$ in this case)  $T\simeq1\%$. In practice $T$ is
usually of the order of $10^{-3}$. It means that the field amplitude $A_0$
changes insignificantly over the coupling zone and can therefore be assumed
constant in the second equation of (\ref{CME}), and the stationary amplitude $A_0\gg B_k$.
Therefore an approximate solution can be obtained:
\begin{eqnarray}
A_0^{out}&=& R A_0^{in}\exp{[i\it \Phi]} + i T_k B^{in}_k \nonumber\\
B_k^{out}&=& B^{in}_k + i T_k A_0,
\label{2fib}
\end{eqnarray}
where
\begin{equation}
T_k = \int\limits_{-\infty}^{\infty}C_k \exp{[i(\beta_0-\beta_k) z]} dz;\;\;\;\;\;\;
\it \Phi = \int\limits_{-\infty}^{\infty}\Delta \beta_0 dz
\end{equation}
Equations (\ref{2fib}) are practically identical to (\ref{Fabry}) if $A_0$
is closed into a ring. In the second equation of (\ref{2fib}) we neglected small
second-order terms while, however, keeping them in the first equation as they
describe the coupler-induced shift in resonant frequency and the reduction of
$Q$ by loading.
\begin{eqnarray}
\omega_0-\omega_0^\prime={{\it\Phi}\over\tau_0}={\omega (n_c^2 - 1)\over 8 \pi \tau_0}{\int\limits_{\bf C} |{\bf e}_0|^2 dv}\nonumber\\
\delta_c={T_k^2\over 2\tau_0}={\omega^2 (n^2_c-1)^2\over 128 \pi^2 \tau_0}\left|{\int\limits_{\bf C} {\bf e}_k^*{\bf e}_0 \exp{[i(\beta_0-\beta_k) z]} dv}\right|^2
\label{fshift}
\end{eqnarray}

\section{VARIATIONAL APPROACH}

Directional coupler approach can be easily generalized for multimode coupler.
However expressions for the coupling parameters better suited for couplers
with dense mode spectrum can be found in a more rigorous way directly from
Maxwell equations using variational methods. Electric field in the resonator
perturbed by coupler may be written in the form:
\begin{equation}
{\bf E}_s({\bf r},t)=\exp{[-i\omega t]}\sum_j \hat A_j(t) {\bf \hat e}_j({\bf r}),
\label{variate}
\end{equation}
where ${\bf \hat e}_j$ are orthonormalized eigenmodes of the unperturbed
lossless resonator without coupler
\begin{equation}
{1\over 4\pi} \int \varepsilon_s{\bf \hat e}_{j1}{\bf \hat e}_{j2}^*dv=\delta_{j1,j2}
\end{equation}
($\delta_{j1,j2}$ is Kronecker symbol here).
Rigorously speaking this normalization meets some
difficulties for open dielectric resonators with finite radiative quality-factor
$Q_{rad}$ {\cite{Lai}}. In our consideration, however, we can avoid them by
assuming eigenfrequences of interest $\omega_j$ to be purely real. For this we neglect the
imaginary part that describes radiation losses and choose as the integration
volume the sphere with a diameter much less than $Q_{rad}\lambda /\pi$.
Amplitudes $\hat A_j$ are slowly varying and differ from circulating
amplitude $A_j$ introduced before only in terms of normalization.
One can easily see that
\begin{equation}
{|{\bf \hat e}_j|^2\over|{\bf e}_j|^2}={c\over 4\pi }{\int\limits_{\bf S}[{\bf \hat e}_j {\bf \hat h}^*_j}] ds ={1\over\tau_j}.
\end{equation}

The equation for the field in the coupled sphere will have the form:
\begin{equation}
\nabla\!\times\!\nabla\!\times\! {\bf E}_s +\left({\varepsilon_s({\bf r})\over c^2}
+{\varepsilon_c({\bf r})-1\over c^2}
+i{2\delta_0\varepsilon_s({\bf r})\over\omega_0 c^2} \right)
{\partial^2{\bf E}_s\over \partial t^2}
= - {\varepsilon_s({\bf r})-1\over c^2}
{\partial^2{\bf E}_c\over \partial t^2},
\label{Maxwell}
\end{equation}
where the second term in brackets is additional polarization due to presence
of the coupler, the third one describes damping associated with intrinsic
losses in the resonator, and the right part is the polarization caused by the
pump wave. Dielectric susceptibilities $\varepsilon_{s|c}({\bf r})$ are
equal to $n_{s|c}^2$ inside and unity outside the spherical resonator and
the coupler correspondingly.
Substituting (\ref{variate}) into (\ref{Maxwell}) and
multiplying this equation by ${\bf \hat e}_0^*$,  after integration over the entire
volume and omitting small terms we obtain:
\begin{equation}
{d\hat A_0\over dt}+\hat A_0(\delta_0+i\Delta\omega^\prime)=
{i\omega(n^2_s-1)\over 8\pi}\exp{[i\omega t]}\int\limits_{\bf S}{{\bf E}_c {\bf \hat e}_0^* dv},
\label{MaxCoupl}
\end{equation}
where $\Delta\omega^\prime=\omega^\prime_0-\omega$ and
\begin{equation}
\omega^\prime_0=\omega_0-{\omega\over 8\pi}\int\limits_{\bf C}{(n^2_c-1)|{\bf \hat e}_0|^2 dv}.
\label{Shift}
\end{equation}
is the new resonance frequency shifted due to the coupler, in total agreement
with (\ref{fshift}).
Let us express the field in the coupler as expansion in travelling modes in
$z$ direction:
\begin{equation}
{\bf E}_c({\bf r},t)= \exp{[-i\omega t]}\int B_{\beta}(z,t) {\bf e}_{\beta}({\bf r})\exp{[i\beta z]} d{\beta},
\label{FieldCoupler}
\end{equation}
Guided localized modes of the coupler in this description can also be easily
taken into account if we choose $B_\beta$ as
\begin{equation}
B_\beta=\sum_k B_k\delta(\beta-\beta_k) + \tilde B_\beta
\end{equation}

The coupler modes are normalized in such a way that
\begin{equation}
{c\over 4\pi}\int [{\bf e}_{\beta1},{\bf h}_{\beta2}^*]_z
ds=\delta(\beta_1-\beta_2)
\end{equation}
(here $\delta(\beta_1-\beta_2)$ is delta-function and ${\bf h}$ is the magnetic
field corresponding to the mode). Integration is performed over
the cross-section orthogonal to $z$-axis. Amplitudes $B_\beta$ (slowly varying
with $z$ and $t$) describe distribution of the pump wave in coupler
modes. Substituting (\ref{FieldCoupler}) into the wave equation:
\begin{equation}
\nabla\!\times\!\nabla\!\times\!{\bf E}_c
+\left({\varepsilon_c({\bf r})\over c^2}+ {\varepsilon_s({\bf
r})-1\over c^2}\right) {\partial^2{\bf E}_c\over \partial t^2} =
-{\varepsilon_c({\bf r})-1\over c^2} {\partial^2{\bf E}_s\over \partial
t^2},
\label{MaxwellC}
\end{equation} we obtain
\begin{eqnarray}
\int{\left(\beta{\partial B_\beta\over \partial z} -
{i\omega^2(\varepsilon_s-1)\over 2 c^2} B_\beta\right){\bf e}_\beta \exp{[i\beta z]} d\beta}
= {{i(\varepsilon_c-1)\omega^2\over2c^2}{\bf E}_s}\exp{[i\omega t]}
\label{ShortC}
\end{eqnarray}
The second term in brackets (\ref{MaxwellC}) determines the change
of the wavenumber (phase velocity) for the given mode in the
coupling zone. Taking vector product of this equation with
${\bf h}_\beta^*$ and integrating over the cross-section, we obtain formal
solutions for slowly varying amplitudes:
\begin{eqnarray}
B_{\beta}&=&B^{in}_{\beta}\exp{[i\Delta\beta z]}+
{i\omega^2\over 8\pi c\beta}\int\limits_{-\infty}^{z} \exp{[i\omega t-i\beta z^\prime
+i\Delta\beta(z-z^\prime)]}\int\limits_{\bf C} {(n^2_c-1)[{\bf E}^\prime_s,
{\bf h}^{\prime*}_\beta]_z ds^\prime  dz^\prime} \nonumber\\
\Delta\beta(z)&=&{\omega^2(n^2_s-1)\over 8\pi c\beta}\int\limits_{\bf S}
[{\bf e}_\beta,{\bf h}_\beta^*]_zds,
\label{Asol}
\end{eqnarray}

Substituting (\ref{FieldCoupler}) into (\ref{MaxCoupl}) using (\ref{Asol})
and omitting $\Delta\beta$, we finally obtain the following equation for
the amplitude of the mode in the resonator:
\begin{equation}
{d\hat A_0\over dt}+(\delta_0+\delta_c+i\Delta\omega^\prime)\hat A_0=
{i\omega (n^2_c-1)\over 8\pi}\int B^{in}_\beta {\int\limits_{\bf C}
{{\bf \hat e}_0 {\bf e}^*_\beta exp{(-i\beta z)} dv} d\beta},
\label{oscfin}
\end{equation}
and
\begin{eqnarray}
\delta_c&=& {\omega^3 \over 64\pi^2 c} \int\!\int\limits_{\bf S}
\!\int\limits_{-\infty}^z\!\int\limits_{\bf C}{(n^2_s-1)(n^2_c-1)\over\beta}
\exp{[i(\beta+\Delta\beta)(z-z^\prime)]}
[{\bf \hat e}^\prime_0,{\bf h}^{\prime*}_\beta]_z ({\bf e}_\beta
{\bf \hat e}^*_0)ds^\prime dz^\prime dv d\beta \nonumber\\
  &\simeq& {\omega^2 (n^2_c-1)^2\over 128 \pi^2}
\int\left | \int\limits_{\bf C}{\bf \hat e}_0{\bf e}^*_\beta \exp{[-i\beta z]} dv\right|^2 d\beta
\label{deltac}
\end{eqnarray}
in natural agreement with (\ref{fshift}).
Total agreement with (\ref{oscillator}-\ref{out1}) becomes apparent if we put
\begin{equation}
T_\beta={\omega (n^2_c-1)\over 8\pi}{\int\limits_{\bf C}
{\bf e}_0 {\bf e}^*_\beta \exp{[i(\beta_0-\beta) z]} dv}
\label{Tb}
\end{equation}

For high-$Q$ WG modes $\beta_0\simeq m/a$, and as the field drops  outside
the resonator approximately as $\exp{[-\gamma r]}$
($\gamma^2\simeq k^2 (n^2_s-1)$), the dependence of
${\bf e}_0$ on $z$ can be approximated as follows:
${\bf e}\simeq {\bf e}(z=0) \exp{[-\gamma z^2/2a]}$.
If the coupler is straight in $z$ direction (as in most demonstrated
couplers to date), we obtain:
\begin{equation}
T_\beta={\omega (n^2_c-1)\over 8\pi}\sqrt{2\pi a\over\gamma}
 \int\limits_{\bf C} \exp{[-(\beta a-m)^2/2\gamma a]}{{\bf e}_0 {\bf e}^*_\beta ds}
\label{Tb2}
\end{equation}

\section{APPLICATION TO DEMONSTRATED COUPLERS }

Let us now use the developed approach for the analysis of coupling of
whispering-gallery modes with optical fiber.
As soon as according to (ref{kcoupl}), possibility of efficient coupling
critically depends on the value of the loading quality-factor $Q_c$ and its
relation with the intrinsic $Q$, in this
section we shall focus on calculation of $Q_c$ and discuss briefly
methods to achieve phase synchronism and mode matching with different
couplers.

To date, two types of optical fiber coupler to WG modes in microsphere
were demonstrated. The first one is the eroded fiber coupler \cite{Arnold1,Arnold2,Eroded},
where evanescent field of a propagating waveguide mode becomes accessible
due to partial removal of the cladding in a bent section of the fiber.
The recently demonstrated second type of the fiber coupler is based on the
stretched section of a single mode fiber employing the mode conversion of
the initial guided wave into waveguide modes of cladding tapered to the
diameter of few microns \cite{taper}.

The most interesting type of strong confinement modes of the sphere - $T\!E_{\ell\ell q}$ with
radius $a$ (where radial index $q$ is small) and the $H\!E_{11}$ mode in the fiber of radius
$b$ can be approximated as follows \cite{resonat,blove}:
\begin{eqnarray}
{\bf \hat e}_s^x&\simeq& {2\sqrt{n^2-1}\over n^2 a^{3/2}}{\left(\ell\over\pi\right)}^{1/4}
\exp{[-\ell(\pi/2-\theta)^2/2 + i\ell\phi]}
\left\{\begin{array}{rl}
{{\rm j}_\ell(knr)/{\rm j}_\ell(kna)} \mbox{ if } r\leq a \\
\exp{[-\gamma(r-a)]} \mbox{ if } r>a
\end{array} \right.\\
{\bf e}_c^x&\simeq& {2\eta\over \gamma b \sqrt{nc}}
\left\{\begin{array}{rl}
{{\rm J}_0(\eta \rho)}/{\rm J}_0(\eta  b) \mbox{ if } \rho\leq b \\
\exp{[-\gamma (\rho-b)]} \mbox{ if } \rho> b
\end{array} \right.
\end{eqnarray}
where
\begin{eqnarray}
\eta b&\simeq&2.405\exp\left(-{1+1/n^2\over 2b\gamma}\right)\simeq 2.405\left(1-{1+1/n^2\over2b\gamma}\right)\nonumber\\
\gamma&\simeq&\sqrt{(\ell+1/2)^2/a^2-k^2} \simeq k\sqrt{n^2-1}
\end{eqnarray}
Using (\ref{fshift}) and several approximations, we can now calculate
\begin{equation}
Q_c\simeq {16\sqrt{2}\pi^5  n^4_s n_c(n_s^2-1)^2\over 9(n_c^2-1)}{a^{3/2}b^3\over\lambda^{9/2}}\exp{[2kd\sqrt{n^2-1}+(\ell-\beta a)^2/\gamma a]},
\label{ql1}
\end{equation}
where $d$ is the gap between the resonator and the fiber and $n_s=n_c=n$.
To obtain optimal coupling, one has to require matching the propagation constants
in the argument of the second exponent of (\ref{ql1}) ($\ell=\beta a$) as in \cite{taper}.
In this case, using approximations for eigenfrequences in the resonator, one can obtain
optimal radius of the fiber and the loaded $Q$.
\begin{eqnarray}
  b\simeq {2.3 a\over\sqrt{(nka)^2-\ell^2}}\simeq 0.51 \left({a\lambda^2\over n^2 (4q-1)}\right)^{1/3}\nonumber\\
Q_c\simeq 102\left({a\over \lambda}\right)^{5/2}{n^3 (n^2-1)\over 4q-1} \exp{[2\sqrt{n^2-1}k d]}
\label{ql2}
\end{eqnarray}

Using the above expressions, we can try to compare our calculations with the
experimental data reported in \cite{taper} for $\lambda=1.55\,\mu m$, $a=85\,\mu m$ and $b=1.7\, \mu m$.
The measured $Q$ was $2\times 10^6$ with $K=72\%$. Using (\ref{kcoupl})
we can obtain $Q_0=8.5\times10^6$ and $Q_c=2.6\times10^6$.
Calculations with (\ref{ql2}) give $Q_c=2.5\times10^6$ -- in agreement
with the experiment.

It is appropriate to note here that in principle, as follows from
(\ref{ql1}), the minimum of $Q_c$ does not correspond to phasematching
($\beta=\ell/a$) and is shifted to smaller $b$. This minimum is also not very
sharp ($\sim2/\sqrt{\ell}$ - several percents of $\beta$). However this case
deserves special consideration, because for smaller $b$ the approximations we
use here will give larger (more than 10\%) error. It is also important that the loaded $Q$
increases very quickly with the size of the resonator (as $\ell^{5/2}$), and in
this way the range of possible applications of such coupler becomes restricted.
Even for very small fused-silica spheres, optimal radius of most common silica fiber
does not correspond to single-mode operation implying further technical
complications in using this type of coupler.

The conclusions of our theory also correlate with the data on limited
efficiency of sidepolished optical fiber couplers
\cite{Arnold1,Arnold2,Eroded}.
Indeed, with the typical monomode fibers having core index equal
or smaller than that of the spheres (made from polystyrene or silica correspondingly in
\cite{Arnold1,Arnold2,Eroded}, with small microspheres
one cannot satisfy phasematching because of relatively large diameter of
standard cores, and with larger spheres  coupling coefficient is too small
(coupling $Q$-factor too high) to provide efficient power insertion into the
resonator.

Efficient coupling (tens of \%) with high-Q microspheres has
been demonstrated with the planar (slab) waveguides \cite{dubr}.
This type of the coupler provides additional freedom compared
to fiber waveguides because it allows free manipulation of the
two-dimensional optical beams. Optimal width for $\ell\ell q$ mode is
$g=2a/\sqrt{2\ell}$. In the meantime requirement of phase matching for
efficient coupling implies optimization of the slab waveguide
thickness $f$. Using the same approach as above, we obtain:

\begin{equation}
Q_c\simeq {8\pi^2 n^3_s n_c (\sqrt{n_s^2-1}+n_s)(n^2_s-1)^{3/2}\over (n^2_c-1)^2}
{af^3\over\lambda^4} \exp{[2kd\sqrt{n^2_s-1}+(\ell-\beta a)^2/\gamma a]}
\label{ql3}
\end{equation}

It is appropriate to note here that either fibers or planar waveguides can
also effectively excite modes with $\ell\neq m$ if the wavevector is inclined
to the ``equator'' plane of the sphere (symmetry plane of the residual ellipticity)
by angle ${\rm arccos}(m/\ell)$. This conclusion becomes evident if we
remember that the mode with $\ell mq$ is equivalent to a precessing inclined
fundamental $\ell\ell q$ mode \cite{precess}.

Prism coupler has been analyzed in our previous papers \cite{prism,precess}
together with precession approach to the description of WG modes and
theoretical and experimental investigation of the far field patterns.
By contrast to waveguide couplers, where practical realization of high
efficiency implies either precise engineering of the waveguide
parameters, or the step-by-step search of the optimal contact point to the
fiber taper, prism coupler allows systematic procedure of coupling
optimization by manipulating the external beams. The two steps to achieve
efficient coupling are 1)adjustment of the incidence angle $\Phi$ of the input
Gaussian beam inside the coupler prism and 2) adjustment of
the angular size of the beam $\Delta \Phi$ and $\Delta \Theta$ to provide
mode matching with far field of the WG mode in the prism ($\it\Gamma$
factor).
\begin{equation}
\sin\Phi_0 = {\ell\over n_c ka}; \;\;
\Delta\Phi^2 = {\sqrt{n^2_s-1}\over n_p^2 ka \cos^2\Phi_0}; \;\;
\Delta\Theta^2 = {n_s+\sqrt{n^2_s-1}\over n_p^2 ka}
\label{ }
\end{equation}
The loading $Q$ with the prism coupler is as follows:

\begin{equation}
Q_c\simeq {\sqrt{2}\pi^{5/2} n_s^{1/2} (n_s^2-1)\over \sqrt{(n^2_c-n^2_s)}}
\left({a\over\lambda}\right)^{3/2} \exp{[2kd\sqrt{n^2_s-1}]}
\label{ql4}
\end{equation}

Fig.2  summarizes the calculations of the loading quality-factor for different types
of couplers (with optimized parameters) in form of the plots of  $Q_c$ under zero gap $d=0$.
The results in Fig.2 allow to quickly evaluate possibility to achieve critical coupling with the
given size and  intrinsic Q of the sphere, along the lines summarized in Sec.2 (\ref{kcoupl}).

In our experiments employing high-$Q$ WG
microsphere resonators, we used prism coupler in most cases and
believe that it remains the most flexible device as it provides ability of
fine adjustment of both phase synchronism and mode matching via convenient
manipulation of the apertures and incidence angles of free beams.
Also, as seen in Fig.2, it provides a significant margin to obtain critical coupling
with the spheres of various size and intrinsic Q. As a result, we routinely obtained coupling
efficiencies to silica microspheres about $30\%$ with standard right-angle BK7
glass prisms — limited by restricted mode overlap due to input refraction
distortions of the symmetrical Gaussian beams. Use of cylindrical
optics or higher refraction prisms to eliminate the mode mismatch
(${\it\Gamma}\to 1$) can significantly improve coupling efficiency and approach
full exchange of energy under critical regime. (About $80\%$ coupling efficiency
is demonstrated further in the experimental Sec.7).

To conclude this section, let us also note here that the critical coupling (that is
characterized by maximal absorption of the input power in the resonator) is in fact
useless for such applications as cavity QED or quantum-nondemolition experiments,
because no recirculated power escapes the cavity mode. To be able to utilize
the recirculated light, one has to provide the inequality $Q_c\ll Q_0$
(strong overcoupling). In other words, the intrinsic quality-factor has to be
high enough to  provide reserve for sufficient loading by the optimal coupler.

\section{PRISM COUPLER: EXPERIMENTAL EFFICIENCY AND
VARIABLE LOADING REGIMES}

In order to illustrate  the results of our analysis, we performed
measurements to characterize coupling efficiency of the prism coupler with
high-Q fused silica microspheres.  As in our previous experiments, we used
microspheres fabricated of high-purity silica preforms by fusion
in a hydrogen-oxygen microburner (see description of the technique in \cite{Therm}).
The focus of present  experiment was to obtain enhanced coupling by
maximizing the mode matching
\begin{equation}
{\it\Gamma}={{\bf T}{\bf B}^{in}\over  T  B^{in}}.
\end{equation}
along with the lines briefly described in  Sec.2.  In our experiment,
to diminish astigmatic distortions of the input gaussian beam at the
entrance face of the coupler prism, we used equilateral prism of flint
glass (SF-18, refraction index n= 1.72). As usual, the input beam
(a symmetrical gaussian beam from a single-mode piezo-tunable He-Ne laser)
was focused onto the inner surface of the prism, at the proximity point
with the microsphere. The angle of incidence and the cross-section of the
input beam were then optimized to obtain maximum response of a chosen
WG mode. Initial alignment was done on the basis of direct observation
of resonance interference in the far field, with the frequency of the
laser slowly swept across the resonance frequency of the mode.  With the
given choice of prism material, optimal angle of incidence for excitation
of whispering-gallery modes $q \simeq 1$  (close to critical angle of
total internal refraction at the silica-glass interface) was approximately equal to
60 degrees so that astigmatic distortions of the input beam at
the entrance face of the prism were minimized.

After preliminary alignment, the coupling efficiency was further maximized
on the basis of direct observation of the resonance absorption dip: full
intensity of the beam after the coupler prism was registered by linear
photodetector and monitored on digital oscilloscope. Results obtained with a
$TM_{\ell\ell q}$ mode (possessing strongest confinement of the field in
meridional direction) are presented in Fig.3 in form of the resonance
curves observed upon successively decreasing coupling (stepwise increasing
gap). Fig.3 illustrates good agreement of theory with experiment: indeed,
resonance transmission decreases with loading until the quality-factor becomes
twice smaller than the intrinsic $Q_0$; after that, intensity contrast of the
resonance decreases. Fig.4 presents explicitly the plot of the fitted
experimental intensity dip versus the loaded quality-factor $Q=Q_c Q_0/(Q_c+Q_0)$,
which yields satisfactory agreement with parabolic prediction from the
generalized expression (\ref{kcoupl}). Maximal contrast of the resonance
obtained in our experiment was $K^2\simeq0.79$ (the "deepest" curve in
Fig.3).

\section{CONCLUSION}

We have presented a general approach to describe the near-field coupling
of high-$Q$ whispering-gallery modes in optical microsphere
resonators to guided or free-space travelling waves in coupler
devices with continuous and discrete spectrum of propagating
modes.

A convenient formalism of the loaded quality-factor to describe
the energy exchange between coupler modes and the resonator
provides a quick algorithm to determine the efficiency of the
given type of the coupler, under given value of the intrinsic
quality-factor of WG modes.

Variable relation between the intrinsic $Q$-factor and loading losses (described
by $Q_c$) through energy escape to coupler modes is a distinctive new property  of
whispering-gallery resonators compared to conventional Fabry-Perot
cavities: the latter are characterized by fixed coupling through the reflectivities of
comprising low-loss mirrors. This unique ability to control the $Q$ and coupling via
WG mode evanescent field allows to obtain new regimes in the devices,
analogous to those available in lumped-element RF and microwave
engineering.

Theoretical estimates on the basis of the suggested theory are in good
agreement with the reported data on the efficiency of different coupler
devices including tapered, sidepolished fiber and slab waveguide.

Original experimental results include direct demonstration of variable loading
and  enhanced efficiency (up to about $80\%$ ) in prism coupler.
Ease of control of phase synchronism and mode overlap between
coupler and microsphere mode by adjusting the input beam parameters make
the prism coupler versatile and efficient for various applications of high-Q
microsphere resonators.

In conclusion, let us note that the near-field coupling may be not a unique method
to efficiently excite highly confined whispering-gallery modes in microspheres.
Simple estimates show that for example, recent advances in optical fiber grating
fabrication methods \cite{Grat} may allow to "imprint" a Bragg-type critical
coupler for high-Q WG modes directly on a sphere made of low-loss germanosilicate
glass. This configuration might be  of special interest for atomic cavity-QED
experiments, where presence of bulky external couplers may destroy the field symmetry,
complicate laser cooling of atoms etc.

\acknowledgments

This research was supported in part by the Russian Foundation for Fundamental
Research grant 96-15-96780.

\vfill
\eject

\begin{figure}
\centerline{\epsfbox{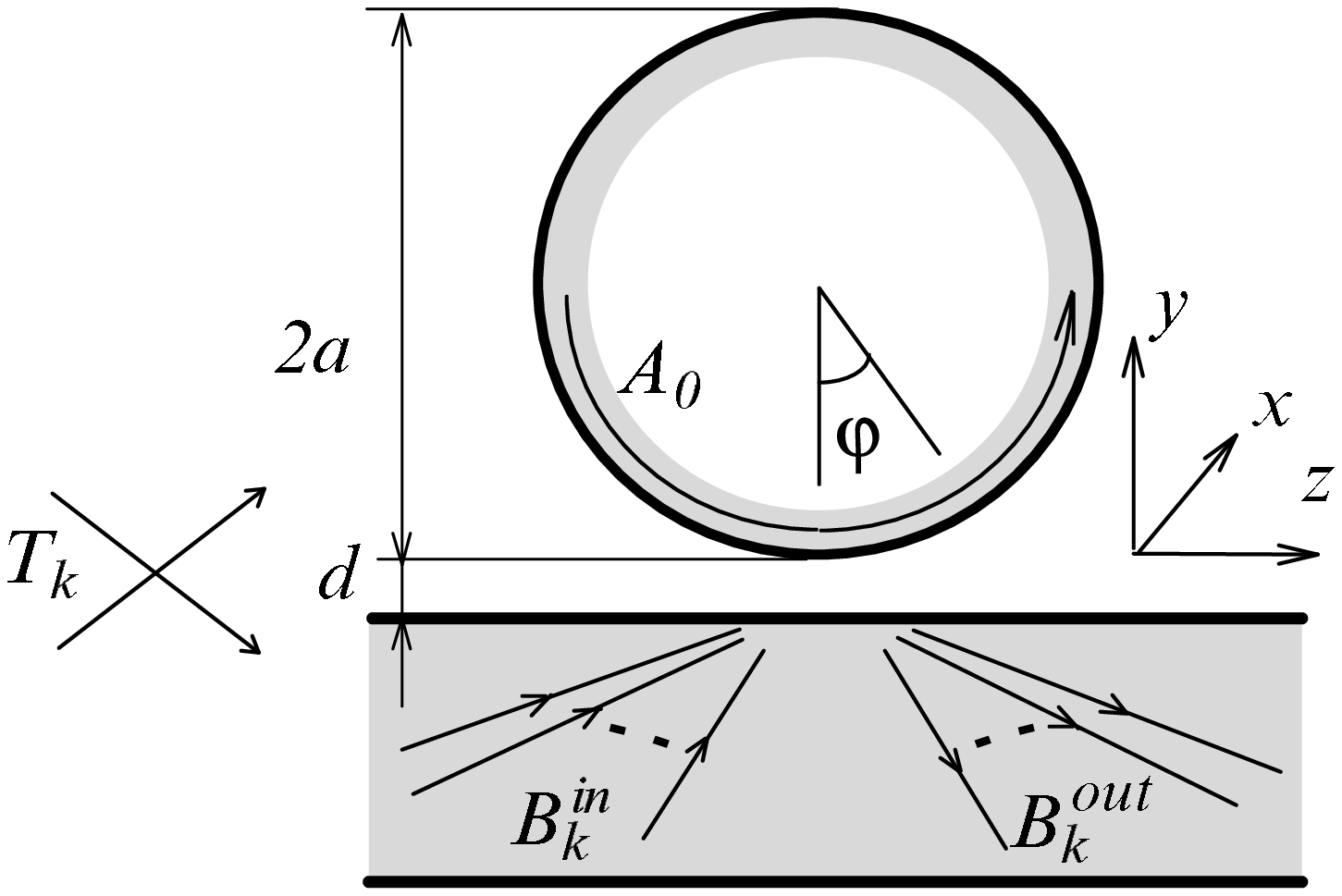}}
\caption
{Schematic of the excitation of whispering gallery modes in high-Q
microsphere}
\label{f1}
\end{figure}
\vfill
\eject

\begin{figure}
\centerline{\epsfbox{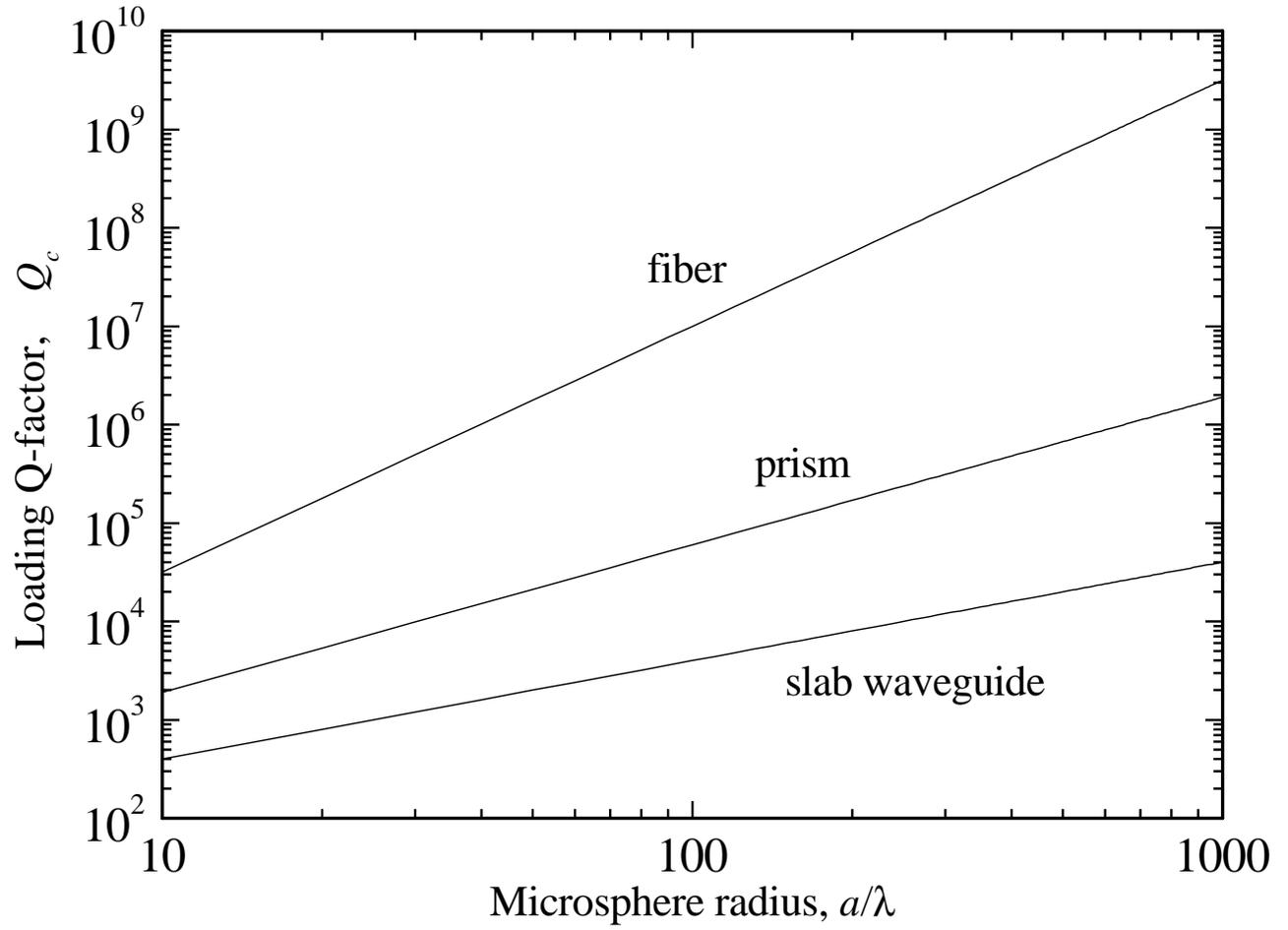}}
\caption
{Efficiency of different couplers in terms of the loading quality-factor
$Q_c$ with optimized parameters at $d=0$ as function of sphere radius $a$;
numerical results are obtained for $TM_{\ell\ell1}$ mode. Critical
coupling is possible when the intrinsic quality-factor of WG mode
$Q_0$ is larger than $Q_c$ (see (\ref{kcoupl}))}
\label{f2}
\end{figure}
\vfill
\eject

\begin{figure}
\centerline{\epsfbox{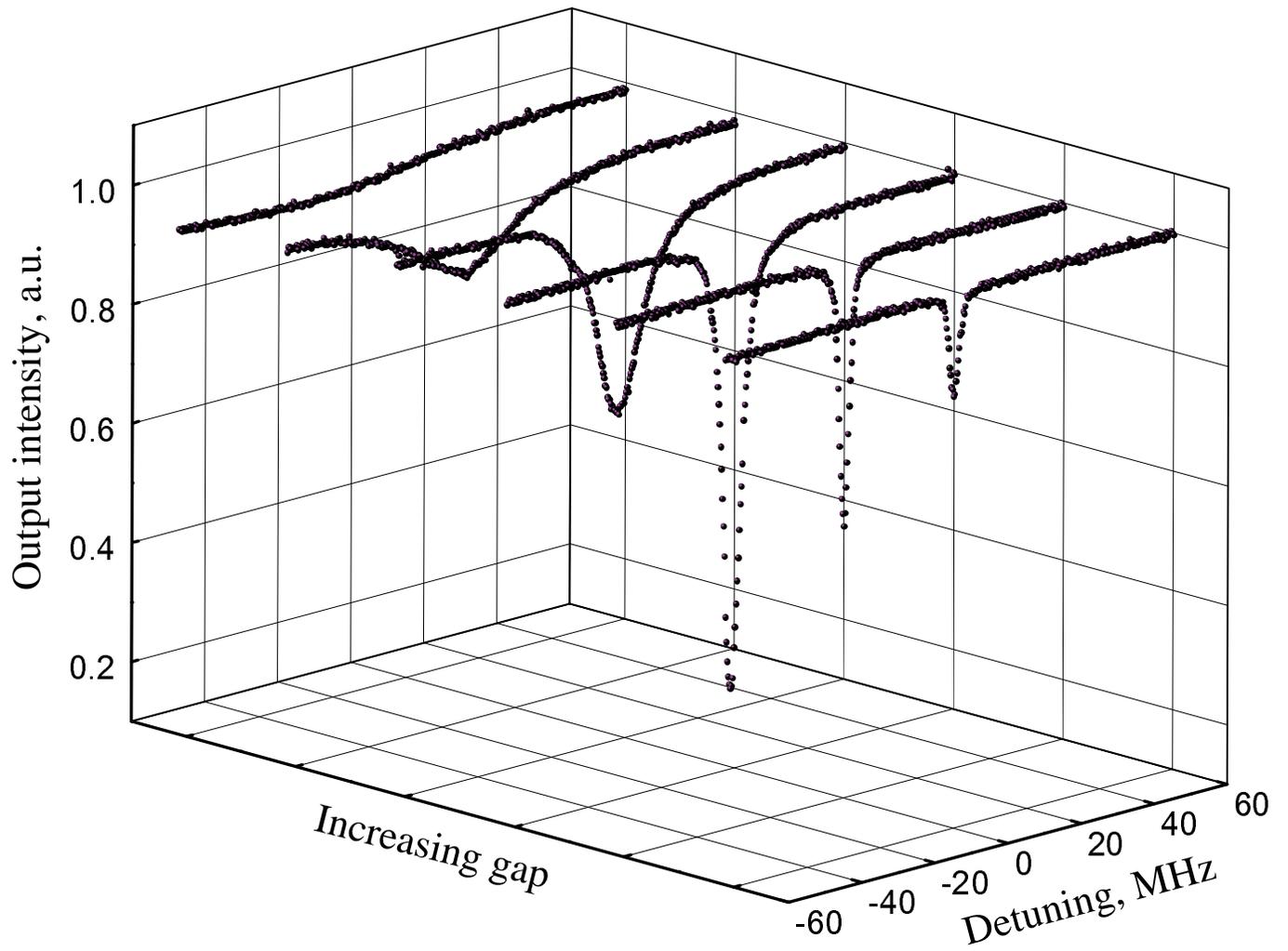}}
\caption
{Output intensity of the prism coupler observed under variable loading
(successively increasing microsphere-prism gap). Fused silica sphere with
diameter $270\mu m$.}
\label{f3}
\end{figure}
\vfill
\eject

\begin{figure}
\centerline{\epsfbox{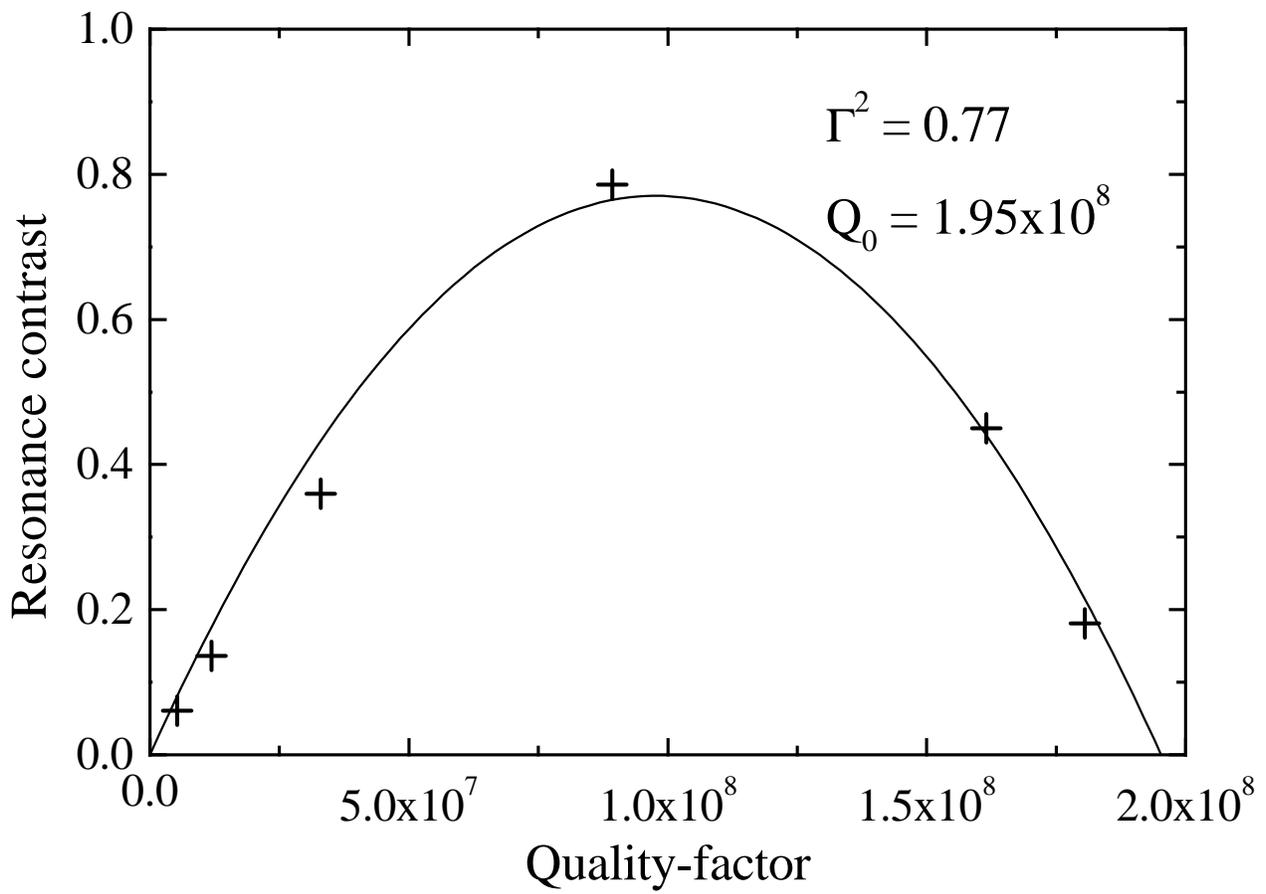}}
\caption
{Resonance contrast as function of the loaded quality factor.}
\label{f4}
\end{figure}
\vfill
\eject

\end{document}